\documentclass[aps,prd,superscriptaddress,amsmath,groupedaddress,twocolumn,floatfix]{revtex4}

\usepackage{color,graphicx,dcolumn,hhline}
\newcommand{\msbar}{\text{$\overline{\text{MS}}$}}
\newcommand{\naive}{\text{na\"\i ve}}
\def\cO{{\cal O}}
\def\chpt{\raise0.4ex\hbox{$\chi$}PT}
\def\schpt{S\raise0.4ex\hbox{$\chi$}PT}
\def\gtwid{$\,$\raise.3ex\hbox{$>$\kern-.75em\lower1ex\hbox{$\sim$}}$\,$}
\def\ltwid{$\,$\raise.3ex\hbox{$<$\kern-.75em\lower1ex\hbox{$\sim$}}$\,$}
\def\MeV{{\rm Me\!V}}
\def\GeV{{\rm Ge\!V}}
\def\eq#1{Eq.~(\ref{eq:#1})}
\def\et{{\it et al.}}

\newcommand{\tmpgreen}{green }
\newcommand{\tmpmagenta}{magenta }
\begin{document}
\title{First determination of the strange and light quark masses from full
lattice QCD}
\affiliation{American Physical Society, One Research Road, Box 9000, Ridge, NY 11961, USA}
\affiliation{Department of Physics, University of Arizona, Tucson, AZ 85721, USA}
\affiliation{Department of Physics, University of California, Santa Barbara, CA 93106, USA}
\affiliation{Laboratory of Elementary-Particle Physics, Cornell University, Ithaca, NY 14853, USA}
\affiliation{School of Physics, University of Edinburgh, Edinburgh, EH9 3JZ, UK}
\affiliation{Department of Physics and Astronomy, University of Glasgow, Glasgow, G12 8QQ, UK}
\affiliation{Department of Physics, Indiana University, Bloomington, IN 47405, USA}
\affiliation{Physics Department, The Ohio State University, Columbus, OH 43210, USA}
\affiliation{Physics Department, University of the Pacific, Stockton, CA 95211, USA}
\affiliation{Physics Department, Simon Fraser University, Vancouver, British Columbia, Canada}
\affiliation{Physics Department, University of Utah, Salt Lake City, UT 84112, USA}
\affiliation{Department of Physics, Washington University, St.~Louis, MO 63130, USA}
\author{${}^{b}$C.~Aubin}\affiliation{Department of Physics, Washington University, St.~Louis, MO 63130, USA}
\author{\ ${}^{b}$C.~Bernard}\affiliation{Department of Physics, Washington University, St.~Louis, MO 63130, USA}
\author{\ ${}^{a,c}$C.~T.~H.~Davies}\affiliation{Department of Physics and Astronomy, University of Glasgow, Glasgow, G12 8QQ, UK}
\author{\ ${}^{b}$C.~DeTar}\affiliation{Physics Department, University of Utah, Salt Lake City, UT 84112, USA}
\author{\ ${}^{b}$Steven~Gottlieb}\affiliation{Department of Physics, Indiana University, Bloomington, IN 47405, USA}
\author{\ ${}^{a,c}$A.~Gray}\affiliation{Physics Department, The Ohio State University, Columbus, OH 43210, USA}
\author{\ ${}^{b}$E.~B.~Gregory}\affiliation{Department of Physics, University of Arizona, Tucson, AZ 85721, USA}
\author{\ ${}^{a,c}$J.~Hein}\affiliation{School of Physics, University of Edinburgh, Edinburgh, EH9 3JZ, UK}
\author{\ ${}^{b}$U.~M.~Heller}\affiliation{American Physical Society, One Research Road, Box 9000, Ridge, NY 11961, USA}
\author{\ ${}^{b}$J.~E.~Hetrick}\affiliation{Physics Department, University of the Pacific, Stockton, CA 95211, USA}
\author{\ ${}^{a}$G.~P.~Lepage}\affiliation{Laboratory of Elementary-Particle Physics, Cornell University, Ithaca, NY 14853, USA}
\author{\ ${}^{a,c}$Q.~Mason}
\affiliation{Laboratory of Elementary-Particle Physics, Cornell University, Ithaca, NY 14853, USA}
\author{\ ${}^{b}$J.~Osborn}\affiliation{Physics Department, University of Utah, Salt Lake City, UT 84112, USA}
\author{\ ${}^{a}$J.~Shigemitsu}\affiliation{Physics Department, The Ohio State University, Columbus, OH 43210, USA}
\author{\ ${}^{b}$R.~Sugar}\affiliation{Department of Physics, University of California, Santa Barbara, CA 93106, USA}
\author{\ ${}^{b}$D.~Toussaint}\affiliation{Department of Physics, University of Arizona, Tucson, AZ 85721, USA}
\author{\ ${}^{a}$H.~Trottier}\affiliation{Physics Department, Simon Fraser University, Vancouver, British Columbia, Canada}
\author{\ ${}^{a}$M.~Wingate}\affiliation{Institute for Nuclear Theory, University of Washington, Seattle, WA 98195, USA}
\collaboration{${}^a$HPQCD, ${}^b$MILC, ${}^c$UKQCD}

\noaffiliation

\date{\today}

\begin{abstract}
We compute the strange quark mass $m_s$ and 
the average of the $u$ and $d$ quark masses $\hat m$ 
using full lattice QCD with three dynamical quarks
combined with the experimental values for the 
$\pi$ and $K$ masses. The simulations have degenerate
$u$ and $d$  quarks with masses $m_u=m_d\equiv\hat m$
as low as $m_s/8$, and two different values of the 
lattice spacing. The bare lattice quark masses
obtained are converted to the $\msbar$ scheme using perturbation 
theory at $\cal{O}$$(\alpha_S)$. Our results are: 
$m_s^\msbar(2\,\GeV)  =  76(0)(3)(7)(0)\;\MeV$,
$\hat m^\msbar(2\,\GeV)  =   2.8(0)(1)(3)(0)\; \MeV$, and
$m_s/\hat m  =  27.4(1)(4)(0)(1)$,
where the errors are from statistics, simulation, perturbation theory,
and electromagnetic effects, respectively. 

\end{abstract}


\maketitle

\section{Introduction}
The masses of the strange 
and light quarks are fundamental parameters
of the Standard Model that are {\it a priori} unknown and 
must be determined from experiment. 
This is 
complicated, however, by confinement in QCD, so that quarks cannot be 
observed as isolated particles. We can only determine their 
masses by solving QCD for observable 
quantities, such as hadron masses, as a function of the quark mass. 
This can be accomplished with the numerical techniques of lattice QCD. 
Precise knowledge of quark masses constrains Beyond 
the Standard Model scenarios as well as providing 
input for phenomenological calculations of Standard 
Model physics.
The strange quark mass, in particular, is needed for various phenomenological 
studies, including the important CP-violating quantity 
$\epsilon^\prime/\epsilon$~\cite{Buras:1996dq}, where 
its uncertainty severely limits the theoretical precision.

Previously, shortcomings in the formulation of 
QCD on the lattice and limitations in computing power have 
meant that lattice calculations were forced to work 
with an unrealistic QCD vacuum that either 
ignored dynamical (sea) quarks or included only 
$u$ and $d$ quarks with masses much heavier than in Nature. 
This condemned determinations of the quark masses 
to rather large systematic errors (10--20\%) arising from the 
inconsistency of comparing such a theory with experiment. 
The determination presented here 
uses simulations with the improved 
staggered quark formalism that have a much more 
realistic QCD vacuum with two light dynamical 
quarks and one strange dynamical quark. 
We describe how the bare quark masses in 
the lattice QCD Lagrangian can be fixed using 
chiral perturbation theory to extrapolate lattice
results to the physical point, and how the lattice quark masses obtained can be 
transformed to a continuum scheme (\msbar) 
using lattice perturbation theory. 
Working in the region of dynamical $u/d$ 
quark masses below $m_s/2$ and down to $m_s/8$ gives us control 
of chiral extrapolations and avoids the 
large systematic errors from dynamical 
quark mass and unquenching effects that previous calculations 
have had.  

Staggered quarks are fast to simulate. They
keep a remnant of chiral symmetry on 
the lattice, and therefore give a
Goldstone pion mass which vanishes with the bare quark mass.
This allows the relatively simple determination of the 
quark mass described here, which is not available, for example, 
in the Wilson quark formalism.

The staggered quark formalism does have several unwanted
features, however.
With the  \naive\ staggered action,
large discretization errors appear, although they 
are formally only $\mathcal{O}(a^2)$ or higher ($a$ is the lattice spacing). The 
renormalization of operators to match a continuum 
scheme can also be large and badly behaved in perturbation 
theory. This is true, for example, for the mass 
renormalization that is needed here. It turns out 
that both problems have the same source, a particular form 
of discretization error in the action, called ``taste violation,'' and 
both are ameliorated by use of 
the improved staggered formalism~\cite{Lepage:1998vj}. 
The perturbation theory then shows small 
renormalizations~\cite{Hein:2001kw,Lee:2002ui,Mason:2002mm} and 
discretization errors are much reduced
~\cite{Blum:1997uf,Bernard:1998mz,Orginos:1998ue}.
Empirically, taste violation remains
the most important discretization error in the improved theory, despite being 
subleading to ``generic'' discretization errors. The Goldstone meson masses
we will discuss here are affected by this 
at one-loop in the chiral expansion. 
Staggered chiral perturbation theory
(\schpt)~\cite{Lee:1999zx,Bernard:2001yj,Aubin:2003mg,Aubin:2003ne}
allows us to control these effects and reduce discretization errors significantly.

A more fundamental concern about staggered fermions is based on the need
to take the fourth root of the quark determinant to convert the four-fold
duplication of ``tastes'' into one quark flavor. It is possible that there are nonlocalities
 in the continuum limit that would spoil the description of QCD at some
level.
Checks of the formalism against experimental results 
\cite{Davies:2003ik,Aubin:2003ne,MILC_SPECTRUM,MILC_FPI,Gottlieb:2003bt},
make this unlikely, we believe, but further work along these lines is 
crucial and continuing. 

\section{Lattice data}\label{sec:data}
The simulation data of the MILC collaboration~\cite{MILC_SPECTRUM,Bernard:2001av} 
are analyzed; staggered quarks with leading errors at 
$\mathcal{O}(\alpha_S a^2,a^4)$~\cite{Lepage:1998vj} and 
one-loop Symanzik improved gluons with 
tadpole-improvement~\cite{Luscher:1985zq,Alford:1995hw}. Two sets of configurations 
are used: a ``coarse'' set at lattice spacing
$a\approx1/8$~fm and sea quark masses of $am'_u=am'_d\equiv a\hat m' =0.005$, 0.007, 0.01, 0.02, 0.03 
with $am'_s=0.05$, and a ``fine'' set at $a\approx1/11$~fm with 
sea quark masses of $a\hat m' =0.0062,$ 0.0124 and $am'_s=0.031$. 
Here we use primes on the sea quark masses to emphasize that
these are the nominal quark masses used in the simulation, not the physical masses
$m_s$ or $\hat m\equiv (m_u +m_d)/2$.
The simulations are ``partially quenched,'' with a range of
valence masses from $m'_s$ down to $m'_s/10$ (coarse) and 
$m'_s/5$ (fine), not necessarily equal to the sea quark masses, simulated
on each lattice.  
It should be noted that the quark masses in lattice units quoted here 
contain a factor of $u_{0P}$, the tadpole-improvement factor determined 
from the fourth root of the average plaquette, 
compared with a more conventional definition of quark mass~\cite{Lepage:1998vj}. 
This is taken care of nonperturbatively before our renormalization below. 

The lattice spacing $a$ is determined ultimately from the 
$\Upsilon^\prime$--$\Upsilon$ mass difference~\cite{agray}, a useful quantity because it is 
approximately independent of quark masses, including the $b$-mass.
An analysis of a wide range of other ``gold-plated'' hadron masses and 
decay constants on these configurations shows agreement with 
experiment at the 2--3\% level~\cite{Davies:2003ik}.
Gold-plated hadrons are stable 
(in QCD), with masses at least $100\; \MeV$ below decay thresholds, 
so their masses are well-defined both experimentally and 
theoretically, important for fixing the parameters of QCD. 
The only gold-plated 
light mesons available to fix $\hat m$ and $m_s$ are 
the $\pi$ and $K$. There is none
with only $s$ valence quarks because the $\phi$ is unstable 
and the pseudoscalar is strongly mixed. Baryons can provide an alternative,
the nucleon for $\hat m$ and the $\Omega$ for $m_s$, but 
their statistical errors are large, and they are not very sensitive to
the quark masses.

Our analysis uses \schpt\ 
\cite{Aubin:2003mg} to fit the 
dependence of the results on the quark masses. 
This dependence can then be extrapolated/interpolated to the point where 
the (Goldstone) $\pi$ and $K$ have their physical masses, thereby
determining the bare lattice $\hat m$ and $m_s$.
At the level of precision at which we are working, and because we 
take $m_u = m_d$, we must be careful about electromagnetic (EM)
and isospin-violating effects.
At lowest nontrivial order in $e^2$ and the quark masses, Dashen's theorem
\cite{Dashen:eg} states that 
$m^2_{\pi^+}$ and $m^2_{K^+}$ receive equal EM
contributions; while the $\pi^0$ and $K^0$ masses are unaffected. However,
at next order, there can be large and different contributions  to $m^2_{\pi^+}$ and $m^2_{K^+}$ of
order $e^2m_K^2$ \cite{Donoghue:hj,Urech:1994hd,Bijnens:1996kk}.   
Let $\Delta_E$ \cite{Nelson-thesis} parameterize violations of Dashen's theorem: 
$(m^2_{K^+}-m^2_{K^0})_{\rm EM}= (1 + \Delta_E)(m^2_{\pi^+}-m^2_{\pi^0})_{\rm EM}$.
Then Refs.~\cite{Donoghue:hj,Urech:1994hd,Bijnens:1996kk} suggest $\Delta_E\approx 1$.

Including EM and isospin effects, the physical values of $\hat m$ and $m_s$ can then be determined by 
extrapolating the lattice squared meson masses to $m^2_{\hat \pi}\equiv m^2_{\pi^0}$ and
$m^2_{\hat K}\equiv (m_{K^0}^2 + m_{K^+}^2 -(1+\Delta_E) ( m_{\pi^+}^2 - m_{\pi^0}^2))/2$, using experimental
values on the right hand side of these expressions. 
We are neglecting $\cO((m_u-m_d)^2)$ corrections, which should be tiny \cite{GASSER_LEUTWYLER}.
EM contributions to the neutral particle masses are also neglected, 
and we take account of this in our error.
For the $\pi^0$ the violation of Dashen's theorem
is $\cO(e^2m_\pi^2/(8\pi^2f_{\pi}^2))$ and negligible. For
$m^2_{K^0}$ the violation is in principle
the same order as for $m_{K^+}^2$ \cite{Urech:1994hd}, but in model
calculations \cite{Bijnens:1996kk} it appears to be
very small. To be conservative, we consider EM contributions to $m^2_{K^0}$ of
order of half the violations of Dashen's theorem, with unknown sign.
Effectively,
this replaces $\Delta_E\approx 1$ in the formula for $m_{\hat K}^2$ above with the
range 0--2, which we take as the EM systematic error.

\section{Chiral Fits and Systematic Errors}

Here we briefly describe the fits to \schpt\ 
theory forms and the estimate of the associated errors~\cite{Aubin:2003ne,MILC_FPI}. 
Because the squared meson
masses ($M^2_{\rm meson}$) are nearly linear in the valence quark masses, 
the final values of the quark masses
are quite insensitive to details of the chiral fits. 
Chiral logs and NLO (and higher)
analytic terms only affect the results at the $\approx 5\%$ level.

\schpt\ is a joint expansion in $x_q$  and $x_{a^2}$, which are dimensionless
measures of the size of quark mass and lattice spacing effects, respectively:
\begin{equation}
x_q \equiv \frac{2\mu m_q}{8\pi^2 f_\pi^2} ; \quad
x_{a^2} \equiv \frac{a^2\overline{\Delta}}{8\pi^2 f_\pi^2} \ .
\end{equation}
$m_q$ is the quark mass, ${2\mu m_q}$ is the tree-level mass of a $q\bar q$ meson,
and $f_\pi\approx 131\; \MeV$.
$a^2\overline{\Delta}$ is an average meson splitting between different tastes. 
On the coarse lattices 
$x_{a^2}\approx 0.09$; on the fine, 
$x_{a^2}\approx 0.03$.  

For physical kaons, the relevant expansion parameter is 
$x_{ud,s}\equiv (x_{ud}+x_s)/2\approx 0.18$.  Since our lattice data is
very precise (0.1 to 0.7\% on $M^2_{\rm meson}$), it is clear that we cannot
expect NLO or even NNLO \chpt\ to work well up to the kaon mass.  If however the
valence quark masses are limited by $m_x+m_y \ltwid 0.75m'_s$, we obtain
good fits including NNLO analytic terms.   Such fits are consistent with \chpt\ expectations:
the coefficients of NLO and NNLO terms are $\cO(1)$  when these terms are expressed as functions of
$x_q$ and $x_{a^2}$.  When fitting up to the strange mass
we include NNNLO as well as NNLO terms, but satisfy the chiral constraints by
fixing the NLO terms from lower mass fits.   Since the $s$ quark mass can be
reached in simulations, the
form of the NNLO and NNNLO terms is not important; 
such terms simply allow for a reasonable interpolation to the physical $m_s$.

Both decay constant and $M^2_{\rm meson}$ data and both coarse and fine 
ensembles are fit simultaneously. 
Although NLO taste-violations are explicitly included, 
we allow for ``generic'' discretization errors  by
using a Bayesian fit~\cite{gplbayes} that permits
physical parameters to change by
order $\alpha_S a^2 \Lambda^2_{QCD}\sim 2\%$ 
in going from the coarse to the fine configurations.

The $\Upsilon$ system provides an absolute lattice scale,
but it is convenient to use the relative scale determined from $r_1$,
a parameter derived from the heavy quark potential~\cite{Sommer:1994ce,Bernard:2000gd},
to compare accurately the scale for different sea quark masses 
within the coarse or fine set.  $\Upsilon$ splittings give
$r_1=0.317(7)(3)\;$fm \cite{MILC_SPECTRUM}.
Using the volume dependence calculated in NLO \schpt\ \cite{Bernard:2001yj,Aubin:2003mg},
(and tested against results on different volumes \cite{MILC_SPECTRUM})
the small finite-volume effects ($<$ 0.75\% in $M^2_{\rm meson}$ ) can be removed from our data with 
negligible residual error. 

\begin{figure}[t]%
\begin{center}%
{\includegraphics[scale=0.35,angle=0,clip]{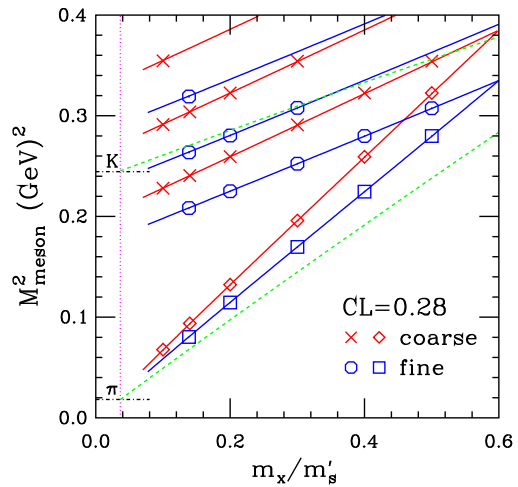}}%
\end{center}%
\caption{Partially quenched data for squared meson masses made out of valence
quarks $x$ and $y$ as a function of
$m_x/m'_s$.
 We show results from two lattices: a coarse
lattice with sea quark masses $a\hat m'=0.01$, $am'_s=0.05$,
and a fine lattice with $a\hat m'=0.0062$,
$am'_s=0.031$.  Three sets of ``kaon'' points
with $m_y=m'_s, 0.8 m'_s,  0.6 m'_s$, are plotted for
each lattice. 
``Pion'' points have $m_x=m_y$.
The solid lines come from a fit to all the data (not just that plotted).
The statistical errors in the points, as
well as the variation in the data with sea quark masses
are not visible on this scale.
The \tmpgreen dashed lines give the continuum fit described in 
the text, and the \tmpmagenta vertical dotted line gives 
the physical $\hat m/m_s$ obtained.} 
\label{F:ChiExtrap}%
\end{figure}

Figure~\ref{F:ChiExtrap} compares our fit with our partially quenched 
data for $M^2_{\rm meson}$.
The data appear quite linear to the eye. 
Indeed, linear fits 
change our result for the quark masses by only $2$ to $7\%$, 
depending on the fit range chosen and whether or not
the correlated decay constants are fit simultaneously. 
However, since the statistical
errors in our data are so small,  the
nonlinearities from chiral logs and higher order analytic terms are crucial for obtaining good fits:
linear fits have $\chi^2/{\rm dof}\! \sim\! 20$.
Nonlinear fits have a confidence level of 0.28, are crucial to obtaining Gasser-Leutwyler parameters and
affect the decay constants by $\sim\! 4$--$12\%$.

We extrapolate/interpolate in mass on the coarse and fine lattices separately to find
the lattice values of the light and strange masses 
that give $m^2_{\hat \pi}$ and $m^2_{\hat K}$.
We get $am_s$ = 0.0390(1)(20), $a\hat m$ = 0.00141(1)(8) on 
the coarse lattices and 0.0272(1)(12) and 0.000989(3)(40) on the 
fine, where errors are statistical and systematic.  The 
systematic errors are dominated by the chiral extrapolation/interpolation, 
estimated by varying the fits,
and the scale uncertainty (EM effects account for the slight difference with~\cite{MILC_SPECTRUM}). 
Alternatively one can
extrapolate the chiral fit parameters to the continuum, 
setting taste-violating parameters zero, and then  
perform the chiral extrapolation/interpolation to the
physical masses.  This is shown as the dashed \tmpgreen
lines in Fig.~\ref{F:ChiExtrap}. 
The methods give final \msbar\ masses that differ by less than 2\%.  We choose the
first method for the central values and include the variation with method in the systematic error.

The same \schpt\ fits that produce the quark masses above give
Gasser-Leutwyler parameters in reasonable 
agreement with phenomenological values~\cite{Aubin:2003ne} and
$f_\pi$ and $f_K$ in agreement with experiment~\cite{Davies:2003ik,Aubin:2003ne}. 
Final results and all details 
of the fits will be described in Ref.~\cite{MILC_FPI}.

\begin{figure}[tb]
\begin{center}%
{\includegraphics[scale=0.35,angle=0,clip]{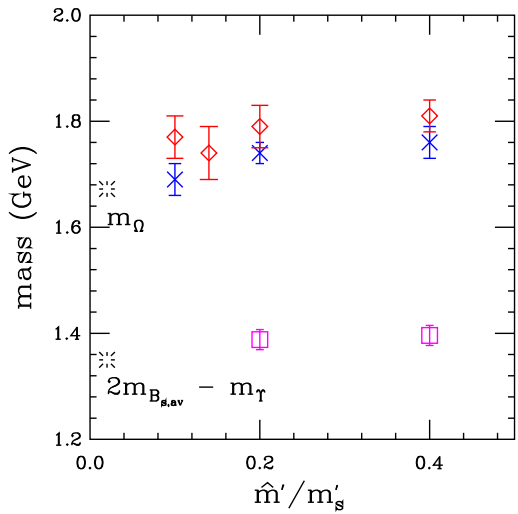}}%
\end{center}%
\caption{Lattice results for two masses which show sensitivity to $m_s$, plotted 
against $\hat m'/m_s'$. The valence $s$ masses are 
taken at the $m_s$ values determined here. The bursts give the corresponding experimental 
result.  
The squares are $2m_{B_{s,av}}-m_\Upsilon$ for two of the
coarse ensembles.
The upper results are for the mass of the $\Omega$ ($sss$) baryon, on both coarse (diamonds) and fine 
(crosses) ensembles. 
\label{F:2mb_mups}}
\end{figure}

It is important to provide further checks of $m_s$ and $\hat m$ using 
other gold-plated masses and mass differences. We focus on
$m_s$ because it has smaller statistical error and less dependence on chiral extrapolations.
From the heavy hadron sector 
$2m_{B_{av},s} - m_{\Upsilon}$ 
is sensitive to $m_s$ but not to other masses. Here 
$2m_{B_{av},s}$ is the $B_s$, $B_s^*$ spin-averaged mass, used
to reduce dependence on the coefficients of 
relativistic corrections in the $b$-quark action. Note, however, that the $B_s^*$ 
is close to decay threshold and
may not be gold-plated.
Figure~\ref{F:2mb_mups} shows coarse-lattice data for this splitting.
The results are 2\% high, but this is also our estimate of discretisation 
errors in the calculation (we do not expect sensitivity to 
taste-violation~\cite{wingate}).  This quantity then provides a check of 
our $m_s$ determination at the 20\% level because the experimental splitting 
varies only by $\approx$ 15\% in changing from $\hat m$ to $m_s$. 
Figure~\ref{F:2mb_mups} also shows results for the $\Omega$ 
baryon mass, on both coarse and fine ensembles. Although 
statistical errors are large there is a trend downwards on 
the finer lattices and signs that a 
continuum extrapolated result will agree with experiment. 
An expected 2\% error on the final value for $m_{\Omega}$ would lead to a 6\% 
determination
of $m_s$.  
\section{Connecting $m^{\text{lattice}}$ with $m^{\msbar}$}\label{sec:perth}
The continuum quark mass in the conventional 
modified Minimal Subtraction scheme is determined from:
\begin{equation}
  m^{\msbar}(\mu)\! = \!\frac{(am)_0}{a} \left(1\!+\!\alpha_V(q^*)\,Z_m^{(2)}\!\left(a\mu,(am)_0\right)
+\mathcal{O}(\alpha^2)\right),
\end{equation}
where $(am)_0$ is the \emph{a posteriori} tuned bare 
mass in lattice units obtained above, converted from the MILC convention by 
dividing by $u_{0P}$.  
$Z_m$ is the mass renormalization that 
connects the bare lattice mass and the $\msbar$ mass. 
The strong coupling constant in the $V$ scheme 
is set using third order perturbative expressions for the 
logarithms of small Wilson loops~\cite{Davies:2002mv,new_alpha} compared 
with lattice results on these configurations. 
The value obtained is run to an optimal scale $q^*$, chosen as described below.  

$Z_m$ is calculated by connecting the bare quark-mass to the pole-mass in 
lattice perturbation theory~\cite{Hein:2001kw}, and using the pole 
mass to $\msbar$ mass relation~\cite{Melnikov:2000qh} at one loop. 
The lattice calculation was done both by hand and using automated 
methods~\cite{Trottier:2003bw,Q:thesis}, which become increasingly 
important for improved actions.  The evaluation
has been checked to lower precision via a completely different method~\cite{Becher:2002if}.
Integrals were evaluated here using the numerical integration package, {\tt VEGAS}~\cite{Lepage:1978sw}. We find
\begin{equation}
\label{eq:Zm}
Z_m^{(2)}(a\mu,am_0) = \left(b(am_0)-\frac4{3\pi}-\frac2\pi\ln(a\mu)\right),
\end{equation}
where $b(am)\approx 0.5432-0.46(am)^2$, correct to 0.1\% up to $(am)=0.1$. 
$\gamma_0=\frac2\pi$ is the universal one-loop anomalous mass dimension. 
Naive staggered quarks have a poorly convergent $Z_m$ with 
$b(0)\approx3.6$ as a result of 
taste-violations. 
It is clear that the improved staggered quark result is much better.  
Tadpole-improvement is also important, because of 
the long paths of gluon fields 
required to suppress taste-violations. 
Without tadpole-improvement $b(0)=2.27$. 

We match our lattice to the \msbar\ scheme at the target scale of
$2\;\GeV$, though the results and errors are not sensitive to this choice.
Because the mass renormalization has 
an anomalous dimension, the optimal $q^*$ value for 
$\alpha_V$ at this scale is dependent on $a$. 
$q^*$ is set by a 
second order BLM method~\cite{Hornbostel:2002af}. On the fine lattices, $q^*$ is 
$1.80/a$~\cite{agray} 
and 
$\alpha_V(q^*)=0.247(4)$ in $Z_m$.  
On the 
coarse lattices, $q^*=2.335/a$, 
giving $\alpha_V(q^*)= 0.252(5)$.
A conservative estimate of the perturbative error in $Z_m$, 
informed by the chiral fits, is
$1.5 \times \alpha_V^2\approx\! 9\%$. 

This gives $m_s^\msbar$ values of 74.3 $\MeV$ on the fine lattices 
and 72.3 $\MeV$ on the coarse lattices. Our central values are obtained by 
extrapolating linearly in $\alpha_S a^2$, the size of the leading discretization errors.
Alternatives, such as a linear extrapolation in $\alpha^2_S a^2$, the size of taste-violations, 
or a continuum-extrapolated chiral fit, give
results that vary by less than 1 MeV, which 
we take as the extrapolation error and fold into the total systematic error. 
Our final quark masses are: 
\begin{eqnarray}
m_s^\msbar(2\,\GeV) & = & 76(0)(3)(7)(0)\; \MeV \\*
\hat m^\msbar(2\,\GeV) & = &  2.8(0)(1)(3)(0)\; \MeV \\*
m_s/\hat m & = & 27.4(1)(4)(0)(1) \ ,
\label{eq:results}
\end{eqnarray}
where the errors come from statistics, simulation systematics, perturbation theory,
and electromagnetic effects, respectively.  
The systematic error includes the scale error in quadrature 
with the chiral and continuum extrapolation errors.
The ratio $m_{s}/\hat m$ in \eq{results} is almost independent of the perturbation 
theory. 
It is also strongly constrained by the fact that $2m_K^2 - m_{\pi}^2$ 
is almost independent of light quark mass over a large 
range. For our coarse lattices it increases by 2\% as $\hat m'$ changes 
from $m_s'/5$ to $m_s'$; for the fine lattices by 4\%. 

\section{Comparison with previous determinations}
There is a long history of sum rule determinations of the strange 
quark mass, with the general trend of decreasing values. 
The current status~\cite{Gamiz:2002nu,Gupta:2001cu,Gupta:2003fn} is broad agreement between 
results from scalar and pseudoscalar spectral functions and 
from SU(3) breaking in $\tau$ hadronic decays, with $m_s$ around 100(20) MeV. The latter 
method, however, is sensitive to the value of $|V_{us}|$. 
Lattice results in the 
quenched approximation give values around 100 MeV but 
more recent results with two flavors of rather heavy dynamical 
quarks give a smaller value around 90 MeV~\cite{Wittig:2002ux}. 
Both quenched and $n_f = 2$ results suffer from the inherent systematic error 
of comparing an unphysical theory with experiment: results 
depend on what hadronic masses are used.
Some determinations also do not use 
gold-plated quantities.  
JLQCD~\cite{Kaneko:2003re} quote a preliminary $n_f$ = 3 result
of 75.6(3.4) MeV, not yet including discretization and finite volume errors.
They use clover quarks with $\hat m'\gtwid m_s/2$, setting 
$a$ with the $\rho$ mass. 

Here we give results from $n_f = 3$ simulations 
in the chiral regime.
Using gold-plated quantities 
to fix the QCD parameters means that there is no remaining 
ambiguity in the match between QCD and experiment.
The value we obtain for $m_s$ is lower than previous results, 
but we maintain that it is based on a firmer footing. 
It violates some quoted bounds from sum rules \cite{Lellouch:1997hp}, 
but these are open to question \cite{Gupta:2003fn}. 
Our result for $m_s/\hat m$ is significantly larger than
that determined from NLO \chpt\ phenomenology \cite{Leutwyler:1996qg}, but
is compatible with a NNLO analysis \cite{Amoros:2001cp}. 
We believe that existing staggered-quark results 
\cite{Davies:2003ik,Aubin:2003ne,MILC_SPECTRUM,MILC_FPI} 
make it unlikely that there are fundamental problems with the formalism we 
are using.
\vspace*{2mm}

\section{Conclusions}
Lattice QCD simulations with improved staggered quarks 
have allowed a new determination of the strange and light quark 
masses with much reduced systematic error:  
our final values are $m_s^\msbar(2\,\GeV)=76(8)\;\MeV$; 
$\hat m^\msbar(2\,\GeV)=2.8(3)\;\MeV$ (adding
errors in quadrature). 
The current lattice simulation error 
can be reduced still further
by generating ensembles with a second (lower) value of the sea strange quark mass and 
is already underway. 
The limiting factor for this determination is 
no longer unquenching but the unknown higher 
order terms in the perturbative mass 
renormalization. 
The two-loop calculation
is clearly needed to improve
our result significantly and is also underway. 
The three-loop errors on masses that would then remain 
would be only $\cal{O}$(2\%), putting the 
determination into a new region of precision. 

This work was supported by the US Department of Energy, the US National Science Foundation, 
PPARC and the EU.
We thank George Fleming, Maarten Golterman, and Rajan Gupta for useful discussions.


\begin{thebibliography}{99}
\bibitem{Buras:1996dq} A.\ J.\ Buras, M.\ Jamin and M.\ E.\ Lautenbacher, Phys.\ Lett.\ {\bf B389}, 749 (1996).
\bibitem{Lepage:1998vj} G.\ P.\ Lepage, Phys.\ Rev.\ {\bf D59}, 074502 (1999).
\bibitem{Hein:2001kw} J.\ Hein \et, Nucl.\ Phys.\ B (Proc.\ Suppl.\ {\bf 106}), 236 (2002); H.\ Trottier \et, ibid, 856. 
\bibitem{Lee:2002ui} W-J.\ Lee and S.\ R.\ Sharpe, Phys.\ Rev.\ {\bf D66}, 114501 (2002).
\bibitem{Mason:2002mm} Q.\ Mason \et\ (HPQCD), Nucl.\ Phys.\ B (Proc.\ Suppl.\ {\bf 119}), 446 (2003).
\bibitem{Blum:1997uf} T.\ Blum \et, Phys.\ Rev.\ {\bf D55}, 1133 (1997).
\bibitem{Bernard:1998mz} C.\ Bernard \et\ (MILC), Phys. Rev. {\bf D58}, 014503 (1998); 
Phys. Rev. {\bf D61}, 111502, (2000). 
\bibitem{Orginos:1998ue} K.\ Orginos and D.\ Toussaint, Phys.\ Rev.\ {\bf D59} 014501 (1999); 
K.\ Orginos, D.\ Toussaint and R.\ L.\ Sugar, Phys.\ Rev.\ {\bf D60} 054503 (1999).
\bibitem{Lee:1999zx} W-J.\ Lee and S.\ R.\ Sharpe, Phys.\ Rev.\ {\bf D60}, 114503 (1999).
\bibitem{Bernard:2001yj} C.\ Bernard (MILC), Phys.\ Rev.\ {\bf D65}, 054031 (2002).
\bibitem{Aubin:2003mg} C.\ Aubin and C.\ Bernard (MILC), Phys.\ Rev.\ {\bf D68}, 034014 (2003); 
ibid, 074011 (2003).
\bibitem{Aubin:2003ne} C.\ Aubin \et\ (MILC), Nucl.\ Phys.\ B (Proc.\ Suppl.\ {\bf 129}), 227 (2004).
\bibitem{Davies:2003ik} C.\ T.\ H.\ Davies \et\ (Fermilab, HPQCD, MILC, UKQCD), Phys.\ Rev.\ Lett.\ {\bf 92} 022001 (2004). 
\bibitem{MILC_SPECTRUM} C.\ Aubin \et\ (MILC), hep-lat/0402030.
\bibitem{MILC_FPI} C.\ Aubin \et\ (MILC), in preparation. 
\bibitem{Gottlieb:2003bt} S.\ Gottlieb, Nucl.\ Phys.\ B (Proc.\ Suppl.\ {\bf 129}), 17 (2004).
\bibitem{Bernard:2001av} C.\ Bernard \et\ Phys.\ Rev.\ D {\bf 64}, 054506 (2001).

\bibitem{Luscher:1985zq} M.\ L\"{u}scher and P.\ Weisz, Phys.\ Lett.\ {\bf B158}, 250 (1985).
\bibitem{Alford:1995hw} M.\ Alford \et, Phys.\ Lett.\ {\bf B361}, 87 (1995).
\bibitem{agray} A.\ Gray \et, Nucl.\ Phys.\ B (Proc.\ Suppl.\ {\bf 119}) 592 (2003). 
\bibitem{Dashen:eg} R.~F.~Dashen, Phys.\ Rev.\  {\bf 183}, 1245 (1969).  
\bibitem{Donoghue:hj} J.~F.~Donoghue, B.~R.~Holstein and D.~Wyler, Phys.\ Rev.\ D {\bf 47}, 2089 (1993).
\bibitem{Urech:1994hd} R.~Urech, Nucl.\ Phys.\ B {\bf 433}, 234 (1995).
\bibitem{Bijnens:1996kk} J.~Bijnens and J.~Prades, Nucl.\ Phys.\ B {\bf 490}, 239 (1997).
\bibitem{Nelson-thesis} D.\ Nelson, hep-lat/0212009.
\bibitem{GASSER_LEUTWYLER} J.~Gasser and H.~Leutwyler, Nucl.\ Phys.\ {\bf B250}, 465 (1985).
\bibitem{gplbayes} G.\ P.\ Lepage \et, Nucl.\ Phys.\ B (Proc.\ Suppl.\ {\bf 106}) 12 (2002). 
\bibitem{Sommer:1994ce} R.\ Sommer, Nucl.\ Phys.\ {\bf B411}, 839 (1994).
\bibitem{Bernard:2000gd} C.\ Bernard \et, Phys.\ Rev.\ {\bf D62}, 034503 (2000). 
\bibitem{wingate} M.\ Wingate \et, Phys.\ Rev.\ D{\bf 67} 054505 (2003). 
\bibitem{Davies:2002mv} C.\ Davies \et, Nucl.\ Phys.\ B (Proc.\ Suppl.\ {\bf 119}) 595 (2003). 
\bibitem{new_alpha} Q.\ Mason, H.\ Trottier \et, in preparation.
\bibitem{Melnikov:2000qh} K.\ Melnikov and T.\ v.\ Ritbergen, Phys.\ Lett.\ {\bf B482}, 99 (2000).
\bibitem{Trottier:2003bw} H.\ D.\ Trottier, Nucl.\ Phys.\ B (Proc.\ Suppl.\ {\bf 129}), 142 (2004).
\bibitem{Q:thesis} Q.\ Mason (2003), Cornell University, PhD thesis. 
\bibitem{Becher:2002if} T.\ Becher and K.\ Melnikov, Phys.\ Rev.\ {\bf D66}, 074508 (2002).
\bibitem{Lepage:1978sw} G.\ P.\ Lepage, J.\ Comput.\ Phys.\ {\bf 27}, 192 (1978).
\bibitem{Hornbostel:2002af} K.\ Hornbostel, G.\ P.\ Lepage and C.\ Morningstar, Phys.\ Rev.\ {\bf D67}, 034023 (2003).
\bibitem{Gamiz:2002nu} E.\ Gamiz \et, JHEP {\bf 01}, 060 (2003).
\bibitem{Gupta:2001cu} R.\ Gupta and K.\ Maltman, Int.\ J.\ Mod.\ Phys.\ {\bf A16S1B}, 591 (2001).
\bibitem{Gupta:2003fn} R.\ Gupta, hep-ph/0311033.
\bibitem{Wittig:2002ux} H.\ Wittig, Nucl.\ Phys.\ B (Proc.\ Suppl.\ {\bf 119}) 59 (2003). 
\bibitem{Kaneko:2003re} T.~Kaneko {\it et al.}  (CP-PACS, JLQCD), Nucl.\ Phys.\ B (Proc.\ Suppl.\ {\bf 129}), 188 (2004). 
\bibitem{Lellouch:1997hp} L.\ Lellouch, E.\ de Rafael and J.\ Taron, Phys.\ Lett.\ {\bf B414}, 195 (1997). 
\bibitem{Leutwyler:1996qg} H.\ Leutwyler, Phys.\ Lett.\ {\bf B378}, 313 (1996).
\bibitem{Amoros:2001cp} G.~Amoros, J.~Bijnens and P.~Talavera, Nucl.\ Phys.\ B {\bf 602}, 87 (2001).
\end{thebibliography}
\end{document}